\documentclass[a4paper,12pt]{article}
\usepackage{graphicx}
\usepackage{jheppub}
\usepackage[T1]{fontenc} % if needed
\usepackage{braket}
\usepackage{graphicx}  % needed for figures
\usepackage{float}
\usepackage{dcolumn}   % needed for some tables
\usepackage{bm}        % for math
\usepackage{amssymb}   % for math
\usepackage{slashed}   % for Dirac Slash
\usepackage{amsmath}   % for mutiline eqn
\usepackage{braket}
\usepackage{simplewick} % for contraction
\usepackage{verbatim}  % for multi-line comment
\usepackage{color} % for textcolor
\usepackage{appendix} % for appendix
\usepackage{subfig}

\title{\boldmath Exact conditions for antiUnruh effect in (1+1)-dimensional spacetime}
\author[a]{Dawei Wu}
\author[b]{Ji-chong Yang}
\author[c,d,a,1]{Yu Shi \note{Corresponding author.}}
\affiliation[a]{Department of Physics \& State Key Laboratory of Surface Physics, Fudan University, Shanghai 200438, China}
\affiliation[b]{Department of Physics, Liaoning Normal University, Dalian 116029, China} 
\affiliation[c]{University of Science and Technology of China, Hefei 230026, China}
\affiliation[d]{Shanghai Research Center for Quantum Science and CAS Center for Excellence in Quantum Information and Quantum Physics, University of Science and Technology of China, Shanghai 201315, China}
\emailAdd{16110190012@fudan.edu.cn}
\emailAdd{yangjichong@fudan.edu.cn}
\emailAdd{yu\_shi@ustc.edu.cn}
\abstract{
Exact conditions for antiUnruh effect in (1+1)-dimensional spacetime are obtained. For detectors with Gaussian switching functions, the analytic results are similar to previous ones, indicating that antiUnruh effect occurs  when the energy gap matches the characteristic time scale. However, this conclusion does not hold for detectors with square wave switching functions, in which case the condition turns out to  depend on both the energy gap and the characteristic time scale in some nontrivial way. We also show analytically that there is no antiUnruh effect for detectors with Gaussian switching functions in (3+1)-dimensional spacetime.}
\begin{document}
\maketitle
\flushbottom
%\tableofcontents

\section{Introduction}

It is well  known~\cite{Unruh:1976db,Davies:1974th,PhysRevD.7.2850} that a uniformly accelerated observer views the Minkowski vacuum as a thermal state with the temperature   proportional to the observer's acceleration $T=a/{2\pi}$,  usually called the Unruh effect. To give a coordinate-invariant characterization of Unruh  effect, people often employ the so-called Unruh-DeWitt detector~\cite{Unruh:1976db,Unruh:1983ms} and study how it ``tinkles'' when accelerated. The simplest Unruh-DeWitt detector is a two-level system, and it is expected that when an accelerating detector     interacts with some quantum field, there is a probability of transition from the initial ground state to the excited state and the probability  increases with the increase of  the acceleration~\cite{Crispino:2007eb}.

However, in recent years it was  found~\cite{Brenna:2015fga} that under some circumstances the transition probability   decreases as  the acceleration increases, which seems to imply that the detector gets cooler when the acceleration increases. This effect is called the antiUnruh effect. Since the antiUnruh effect is defined according to the behavior of detectors, it is, unlike the Unruh effect, highly dependent on the types of detectors.

The antiUnruh effect may lead to the enhancement of the entanglement between Unruh-DeWitt detectors~\cite{Li:2018xil,Foo:2021gkl,Zhou:2021nyv,Chen:2021evr}. Moreover, the results can be applied to black holes~\cite{Henderson:2019uqo,DeSouzaCampos:2020ddx,deSouzaCampos:2020bnj,Robbins:2021ion,Conroy:2021aow} and other thermal systems~\cite{Pan:2021nka,Barman:2021oum}. However, despite some discussions on the mechanism of antiUnruh phenomena~\cite{Brenna:2015fga,PhysRevD.94.104048}, the physical reason for it remains unclear.

In this paper, we derive the exact conditions of the antiUnruh effect  for detectors with Gaussian and square wave switching functions. In (1+1)-dimensional spacetime,  for Gaussian switching functions, the antiUnruh effect  appears when  $\Omega \sigma < 1/\sqrt{2}$ while for square wave switching functions, the antiUnruh effect  appears when $\left(2 (\Omega \sigma)^2-3\right) \cos (2 \Omega \sigma)-4 \Omega \sigma \sin (2 \Omega \sigma)+3<0$, where   $\Omega$ and $\sigma$ are the energy gap and the characteristic switching time respectively. We also find that  no antiUnruh effect exists  in (3+1)-dimensional spacetime, at least for Gaussian switching functions. We expect our analytic calculations and results   be useful in revealing the physical reason of the antiUnruh effect .

This paper is organized as the following. In Section II we review the basic model for the antiUnruh effect  in (1+1)-dimensional and (3+1)-dimensional spacetimes. We present and analyze our main results in Section III. Section IV is the summary and conclusion.

\section{Model}
In this section, we review the simplest model for this effect~\cite{Brenna:2015fga}. First, we consider a uniformly accelerated two-level Unruh-DeWitt detector with the energy gap $\Omega$ in (1+1)-dimensional Minkowski spacetime. The detector interacts with a massive scalar field $\varphi$, with the interaction Hamiltonian
\begin{equation}
\begin{aligned}
H_{I}=\lambda\chi(\tau,\sigma)\mu(\tau)
\varphi\left(x\left(\tau\right),
t\left(\tau\right)\right),
\end{aligned}
\end{equation}
where $\lambda$ is the strength of the coupling and $\tau$ is the proper time along the detector's worldline, $\mu(\tau)=\exp{(i\Omega\tau) }\sigma^{+}+\exp{(-i\Omega\tau )}\sigma^{-}$ is the monopole operator, $\chi$ is the switching function, which we can, for example, choose as the Gaussian type
\begin{equation}
\begin{aligned}
\chi(\tau,\sigma)=e^{-\frac{\tau^2}{2\sigma^2}},
\end{aligned}
\end{equation}
with $\sigma$ being the characteristic time.

Suppose the initial state is $\ket{g}\ket{0}$, where $\ket{g}$ refers to the ground state of the detector and $\ket{0}$ refers to the vacuum state of the scalar field in the Minkowski spacetime. The evolution of the system is
\begin{equation}
\begin{aligned}
U\ket{g}\ket{0}=\left(1-i\int d \tau H(\tau)+\cdots\right)\ket{g}\ket{0}
\end{aligned}
\end{equation}
where we have used the perturbation expansion. Given the monopole operator $\mu(\tau)=\exp{(i\Omega\tau) }\sigma^{+}+\exp{(-i\Omega\tau )}\sigma^{-}$ and the mode expansion of the massive scalar field in (1+1)-dimensional Minkowski spacetime $\varphi(x,t)=\int \frac{dk}{\sqrt{4\pi \omega}}\left[a(k)e^{-i(\omega t-kx)}+a^\dagger (k)e^{i(\omega t-kx)}\right]$, where $\omega=\sqrt{k^2+m^2}$, $m$ is the mass of the scalar field, we obtain the final state of the system, 
\begin{equation}
\begin{aligned}
\ket{g}\ket{0}-i\lambda\int d\tau \chi(\tau,\sigma)e^{i\Omega \tau}\int \frac{dk}{\sqrt{4\pi \omega}} e^{i(\omega t-kx)} \ket{e}\ket{1}_{k}
\end{aligned}
\end{equation}
where $\ket{e}$ is the excited state of the detector and $\ket{1}_{k}$ is the one-particle state of the field in mode $k$. The typical trajectory of a uniformly accelerated detector can be given as  $x(\tau)=a^{-1}\left(
\cosh{\left(a\tau\right)}-1\right)$ and $t(\tau)=a^{-1}\sinh{(a\tau)}$, where $a$ is the acceleration. Therefore the transition probability is
\begin{equation}
\begin{aligned}
P(\Omega,a,\sigma,m)=\int_{-\infty}^{+\infty} dk |I_k|^2,
\end{aligned}
\end{equation}
with
\begin{equation}
\begin{aligned}
I_k&=\frac{\lambda}{\sqrt{4\pi \omega}}\int_{-\infty}^{+\infty}d\tau \chi (\tau,\sigma) \exp\left(i\Omega \tau \right.\\
&\left.+i\frac{\omega}{a}\sinh{a\tau}-i\frac{k}{a}\left(\cosh{a\tau}-1\right)\right).
\end{aligned}
\end{equation}

Similar results hold for antiUnruh effect  in (3+1)-dimensional  spacetime. In such a case, the scalar field mode expansion is
\begin{equation}
\begin{aligned}
\varphi(\vec x,t)=\int \frac{d^3 \vec{k}}{\sqrt{(2\pi)^3 2 \omega}}\left[a(\vec{k})e^{-i(\omega t-\vec{k}\cdot \vec x)}+a^\dagger (\vec{k})e^{i(\omega t-\vec{k} \cdot \vec x)}\right] .
\end{aligned}
\end{equation}
Suppose the detector accelerates along  x axis  ($y=z=0$). Then following similar calculations, we obtain the final state
\begin{equation}
\begin{aligned}
\ket{g}\ket{0}-i\lambda\int d\tau \chi(\tau,\sigma)e^{i\Omega \tau}\int \frac{d^3 \vec{k}}{\sqrt{(2\pi)^3 2 \omega}} e^{i(\omega t-\vec{k} \cdot \vec x)} \ket{e}\ket{1}_{\vec k},
\end{aligned}
\end{equation}
and the transition probability
\begin{equation}
\begin{aligned}
P(\Omega,a,\sigma,m)=\int_{-\infty}^{+\infty} d^3 \vec{k} |I_{\vec{k}}|^2,
\end{aligned}
\end{equation}
with
\begin{equation}
\begin{split}
I_{\vec{k}}&=\frac{\lambda}{\sqrt{(2\pi)^3 2 \omega}}\int_{-\infty}^{+\infty}d\tau \chi (\tau,\sigma) \exp\left(i\Omega \tau \right.\\
&\left.+i\frac{\omega}{a}\sinh{a\tau}-i\frac{k_x}{a}\left(\cosh{a\tau}-1\right)\right).
\end{split}
\end{equation}

\section{Results}
In this section, we present the analytic conditions for antiUnruh effects  in (1+1)-dimensional and (3+1)-dimensional spacetimes. The details of the calculation are given in the Appendix.

\subsection{(1+1)-dimensional spacetime}
In the case of $D=1+1$, we focus on Unruh-DeWitt detectors with Gaussian or square wave switching functions, which can be written as
\begin{equation}
\begin{split}
\chi^{(G)} (\tau, \sigma)&=\frac{1}{\sqrt{2\pi}\sigma}e^{-\frac{\tau ^2}{2\sigma^2}},\\
\chi ^{(S)}(\tau, \sigma)&=\frac{1}{2\sigma}H(\sigma - \tau)H(\sigma + \tau),
\end{split}
\end{equation}
where $H$ is the Heaviside step function. Note that the Fourier transformations of the switching functions are
\begin{equation}
\begin{split}
\tilde{\chi}^{(G)} (\omega, \sigma)&= \frac{e^{-\frac{\omega^2\sigma^2}{2}}}{\sqrt{2\pi}},\\
\tilde{\chi}^{(S)} (\omega, \sigma)&= \frac{\sin (\sigma \omega)}{\sqrt{2\pi}\sigma \omega},\label{Frou}
\end{split}
\end{equation}
and they are both square integrable,
\begin{equation}
\begin{split}
\int _{-\infty}^{\infty}d\omega\left|\tilde{\chi}^{(G)} (\omega, \sigma)\right|^2&=\frac{1}{2\sqrt{\pi}\sigma},\\
\int _{-\infty}^{\infty}d\omega\left|\tilde{\chi}^{(S)} (\omega, \sigma)\right|^2&=\frac{1}{2\sigma}.
\end{split}
\end{equation}

\subsubsection{Gaussian switching function}
We start with Gaussian switching functions. As shown in Eqs.~(\ref{eq.ap1.17}) and (\ref{eq.ap1.22}), we obtain the analytic expression for the transition probability in the small mass limit,

\begin{equation}
\begin{split}
&P_{\pm}^{(G)}=\frac{1}{2\pi \sigma ^2}\left(P_{LO}^{(G)}+P_{NLO}^{(G)}\right)+\mathcal{O}(m^3),\\
&P_{LO}^{(G)}=\sigma ^2e^{-\Omega ^2 \sigma ^2}\left\{\left(\Omega ^2\sigma^2 \;_2F_2\left(\left.\begin{array}{c} 1, 1\\ \frac{3}{2}, 2\end{array}\right|\Omega^2\sigma^2\right)-\log \frac{m\sigma}{2}-\frac{\pi}{2} {\rm erfi}(\Omega \sigma)-\frac{\gamma _E}{2}\right)\right.\\
&\left.-\frac{a^2\sigma^2}{12}\left(1-2\Omega^2\sigma^2-\frac{a^2\sigma^2}{60}\left(4\Omega^4\sigma^4-12\Omega^2\sigma^2+3\right)\right)\right\}+\mathcal{O}(a^6\sigma^8),\\
&P_{NLO}^{(G)}=\frac{2m^2\sigma^4}{\sqrt{\pi}}\left(2 \Omega\sigma \left.\left(\frac{d}{dx}\; _1F_1\left(\left.\begin{array}{c} x \\ \frac{3}{2}\end{array}\right|-\Omega^2\sigma^2\right)\right) \right|_{x=2}\right.\\
&\left.-\left(\left(2 \Omega^2\sigma^2-1\right) F(\Omega \sigma)-\Omega \sigma\right) (2 \log (m \sigma)+\pi +\gamma _E -1)\right),\\
\end{split}
\end{equation}
where $_pF_q$ are generalized hypergeometric function, as defined in Eq.~(\ref{eq.ap1.6}). Note that although infrared divergence is encountered in $\log (m\sigma)$, the result is still valid for small mass.

\begin{figure*}
\subfloat[$\Omega = 0.1$ and $\Omega \sigma < 1/\sqrt{2}$]{\includegraphics[width=0.49\textwidth]{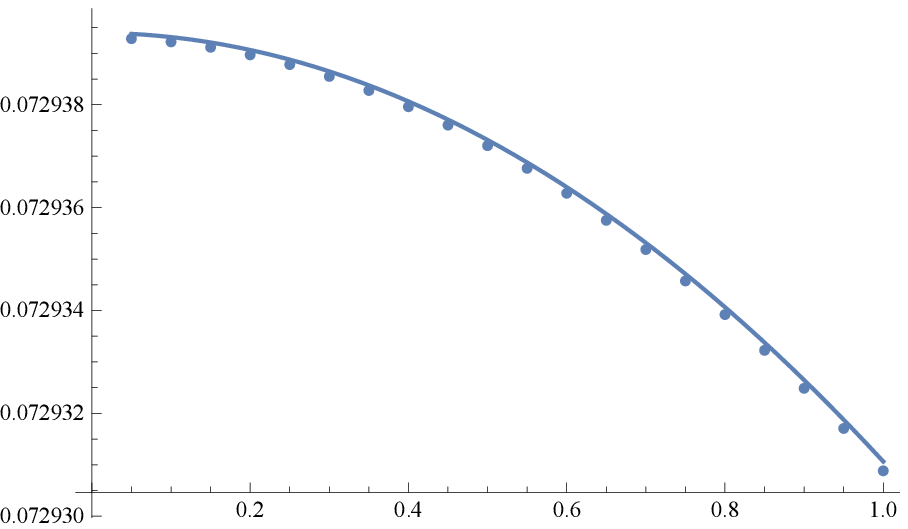}}\hfill
\subfloat[$\Omega = 2$ and $\Omega \sigma < 1/\sqrt{2}$]{\includegraphics[width=0.49\textwidth]{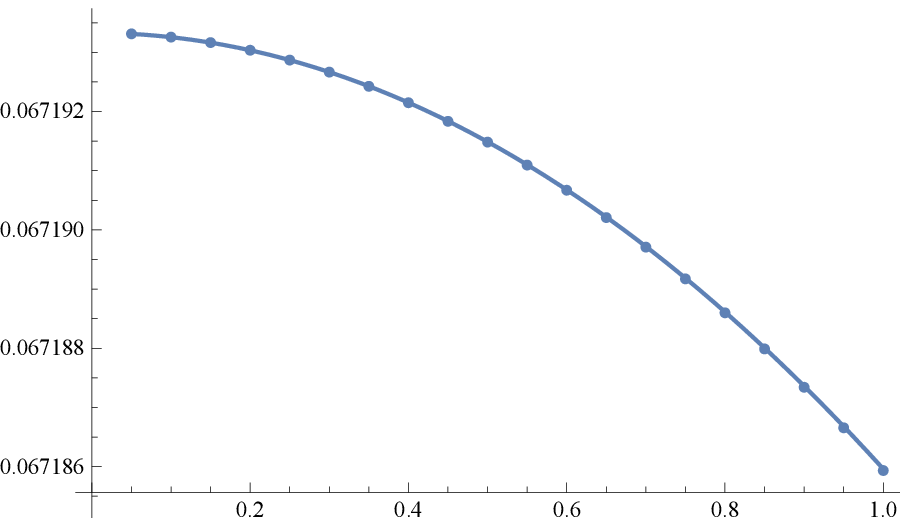}}\vfill
\subfloat[$\Omega = 8$ and $\Omega \sigma > 1/\sqrt{2}$]{\includegraphics[width=0.49\textwidth]{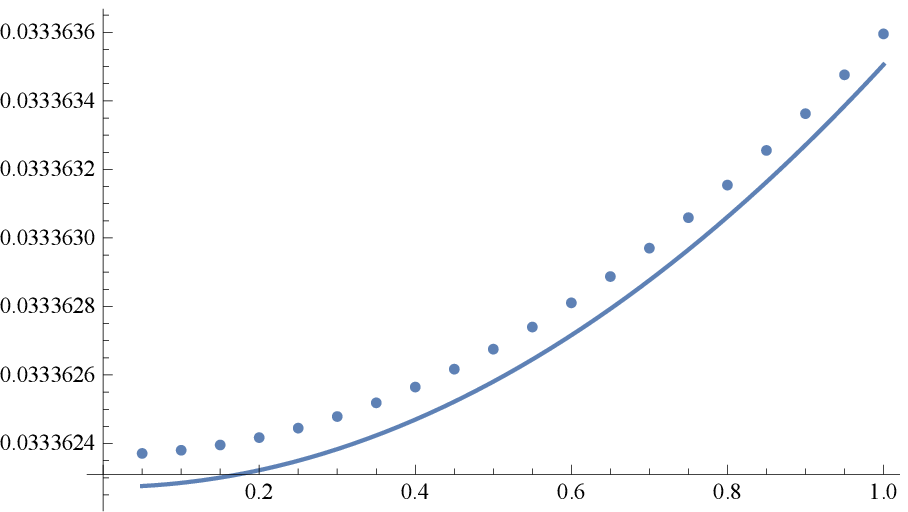}}\hfill
\subfloat[$\Omega = 10$ and $\Omega \sigma > 1/\sqrt{2}$]{\includegraphics[width=0.49\textwidth]{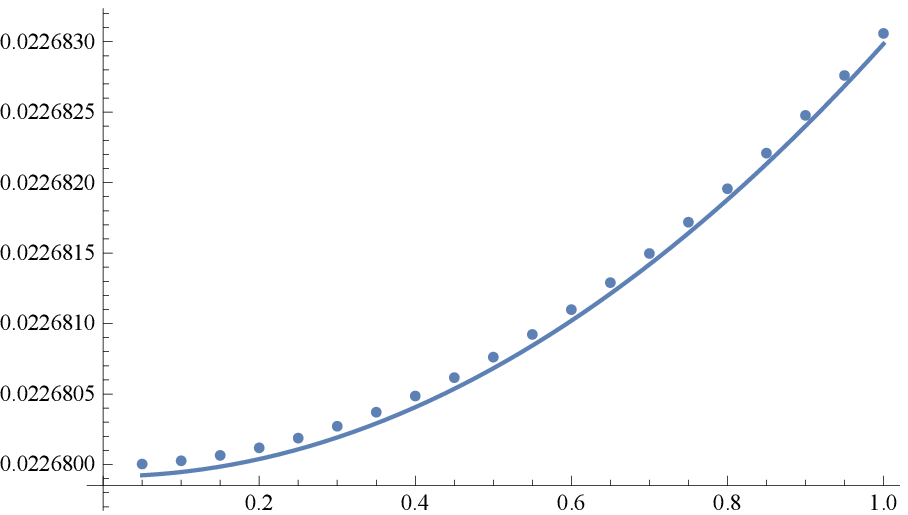}}\vfill
\caption{Comparison of the numerical results $2\pi \sigma ^2 P_{\pm}$ and the analytic results $P_{LO}^{(G)}+P_{NLO}^{(G)}$ when $\sigma=0.1$ and $m=0.01$. We performed numerical integrals with cutoff $\int _{-30}^{30}dk \left|\int _{-1}^1 d\tau I(\tau,k)\right|^2$.}
\label{fig:res1}
\end{figure*}

\begin{figure*}
\subfloat[$\Omega = 0.1$ and $\Omega \sigma < 1/\sqrt{2}$]{\includegraphics[width=0.49\textwidth]{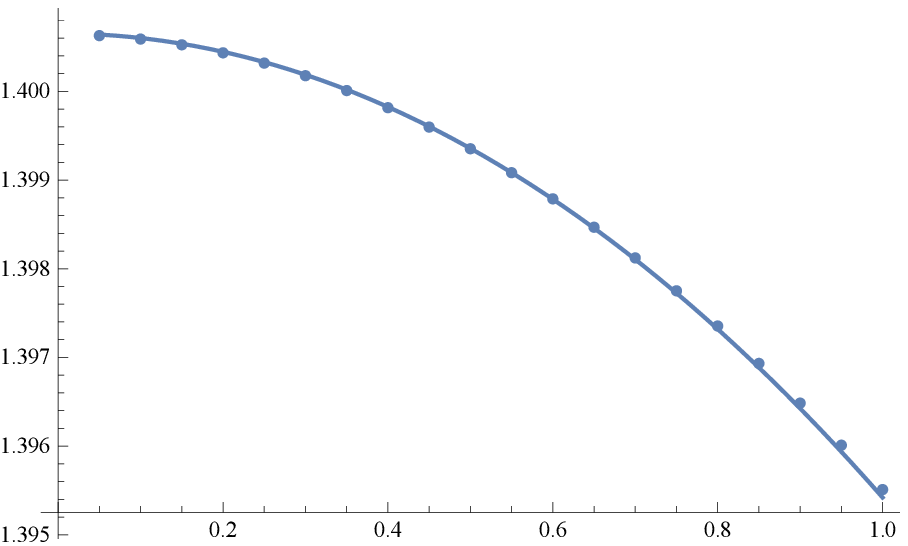}}\hfill
\subfloat[$\Omega = 0.5$ and $\Omega \sigma < 1/\sqrt{2}$]{\includegraphics[width=0.49\textwidth]{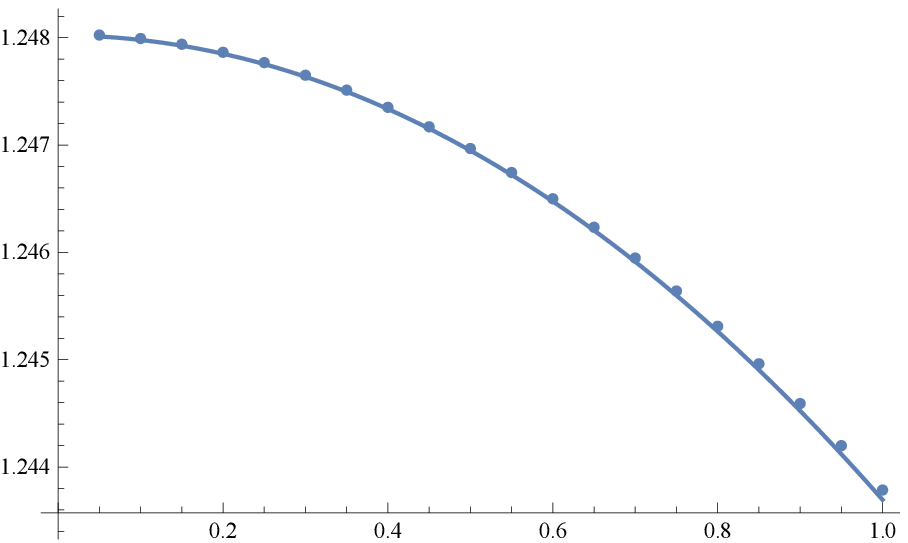}}\vfill
\subfloat[$\Omega = 2$ and $\Omega \sigma > 1/\sqrt{2}$]{\includegraphics[width=0.49\textwidth]{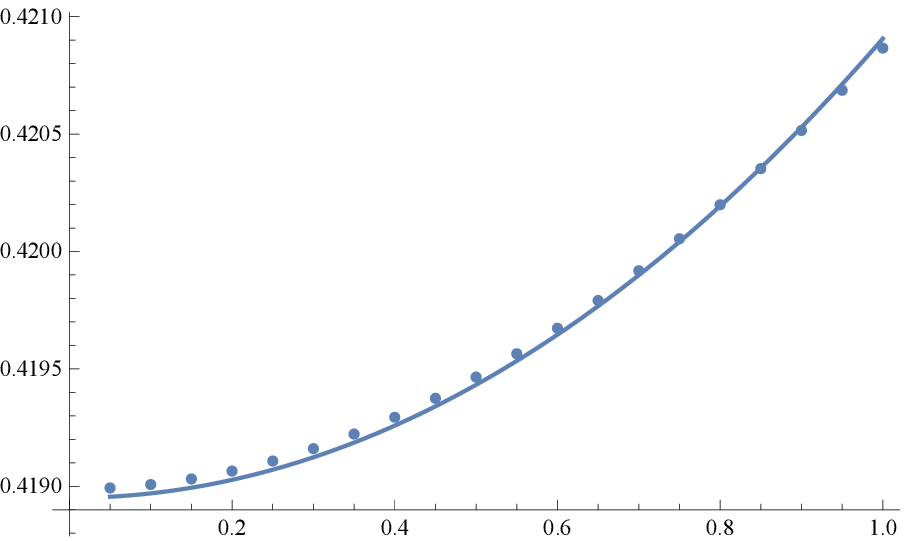}}\hfill
\subfloat[$\Omega = 5$ and $\Omega \sigma > 1/\sqrt{2}$]{\includegraphics[width=0.49\textwidth]{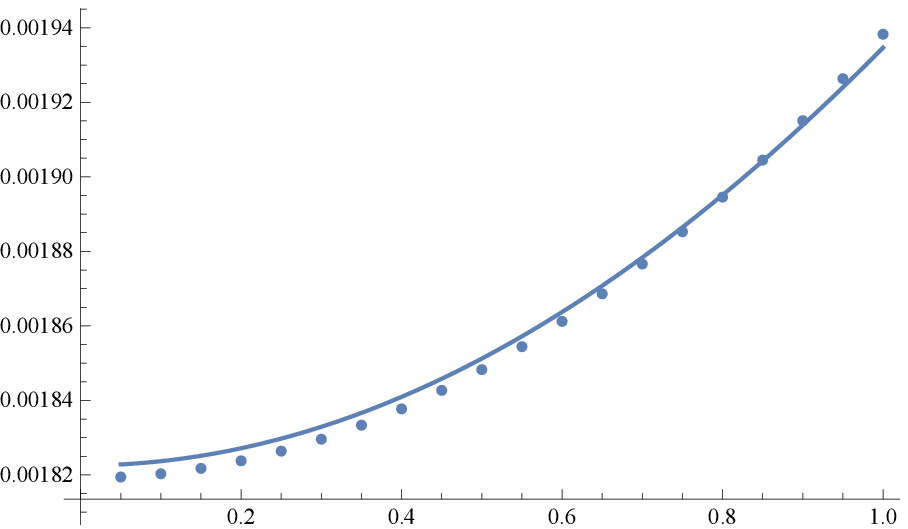}}\vfill
\caption{Comparison of the numerical results $2\pi \sigma ^2 P_{\pm}$ and the analytic results $P_{LO}^{(G)}+P_{NLO}^{(G)}$ when $\sigma=0.5$ and $m=0.01$. We performed numerical integrals with cutoff $\int _{-10}^{10}dk \left|\int _{-6}^6 d\tau I(\tau,k)\right|^2$.}
\label{fig:res2}
\end{figure*}

At the leading order of $a^2$, the coefficient of $a^2$ is $1-2\Omega^2\sigma^2$. Therefore the antiUnruh effect can be found at
\begin{equation}
\begin{split}
&\Omega \sigma < \frac{1}{\sqrt{2}}.\\
\end{split}
\end{equation}
This is in agreement with the original statement of antiUnruh effect  which claims the interaction time interval to be finite $\sigma \sim \Omega^{-1}$~\cite{Brenna:2015fga}. The comparison of the analytic and numerical results is shown in Figs.~\ref{fig:res1} and \ref{fig:res2}.

\subsubsection{Square wave switching function}
\begin{figure*}
\subfloat[$\Omega = \pi / 2$, $\sigma =2$ and $\Omega \sigma =\pi$]{\includegraphics[width=0.49\textwidth]{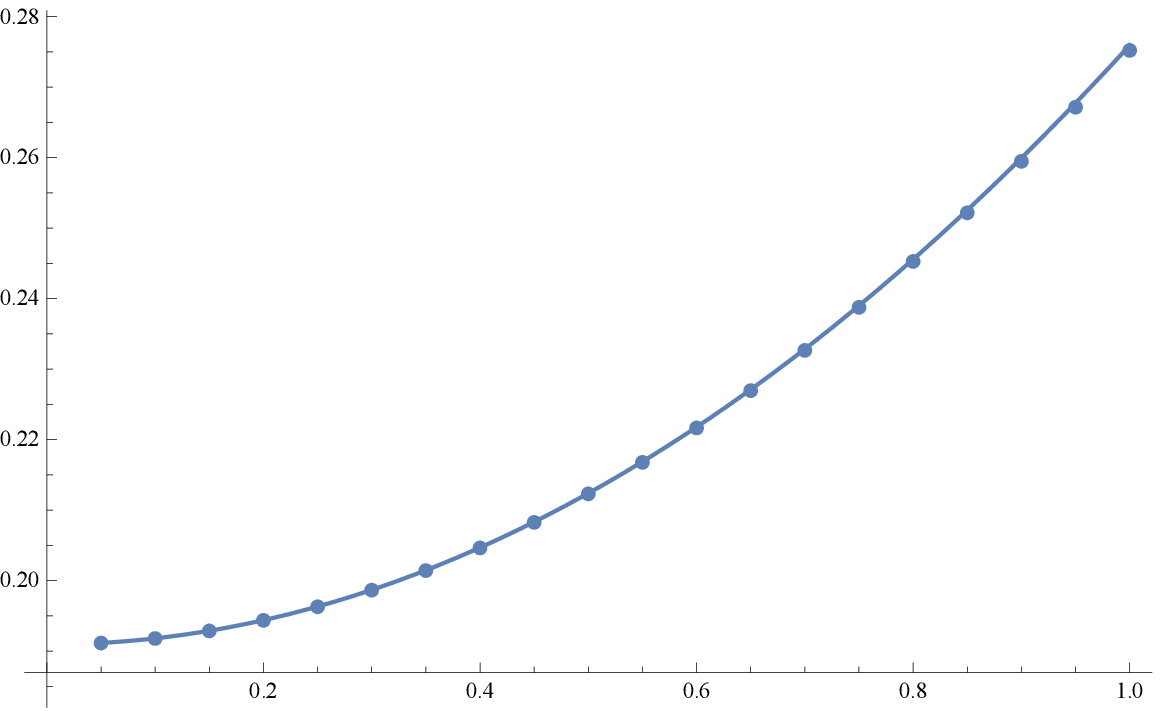}}\hfill
\subfloat[$\Omega = 3\pi / 4$, $\sigma =2$ and $\Omega \sigma =1.5\pi$]{\includegraphics[width=0.49\textwidth]{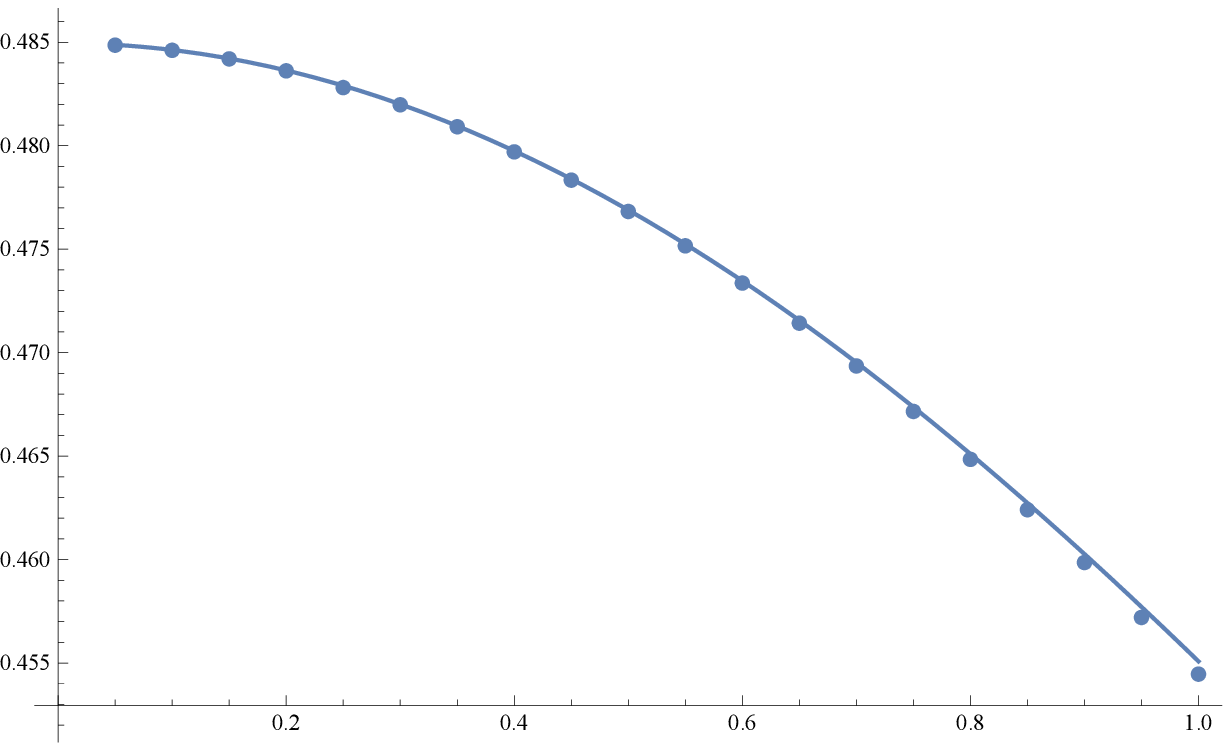}}\vfill
\subfloat[$\Omega = 10$, $\sigma =2\pi$ and $\Omega \sigma =20\pi$]{\includegraphics[width=0.49\textwidth]{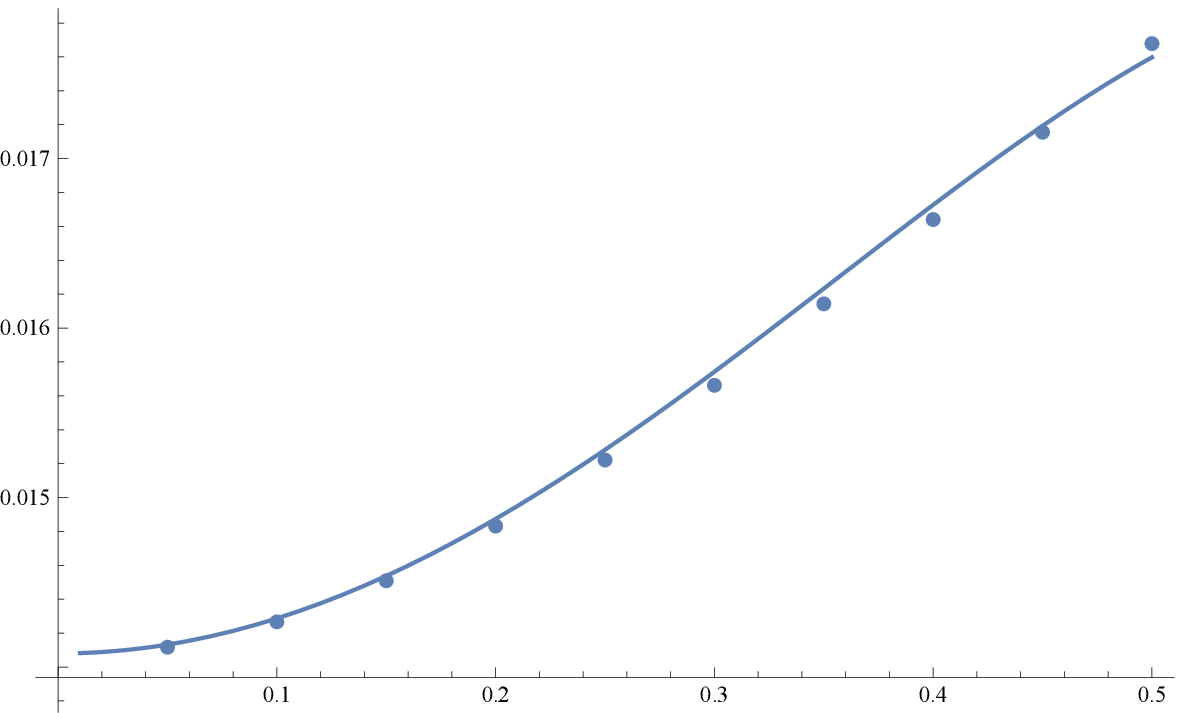}}\hfill
\subfloat[$\Omega = 10$, $\sigma =41\pi/20$ and $\Omega \sigma =20.5\pi$]{\includegraphics[width=0.49\textwidth]{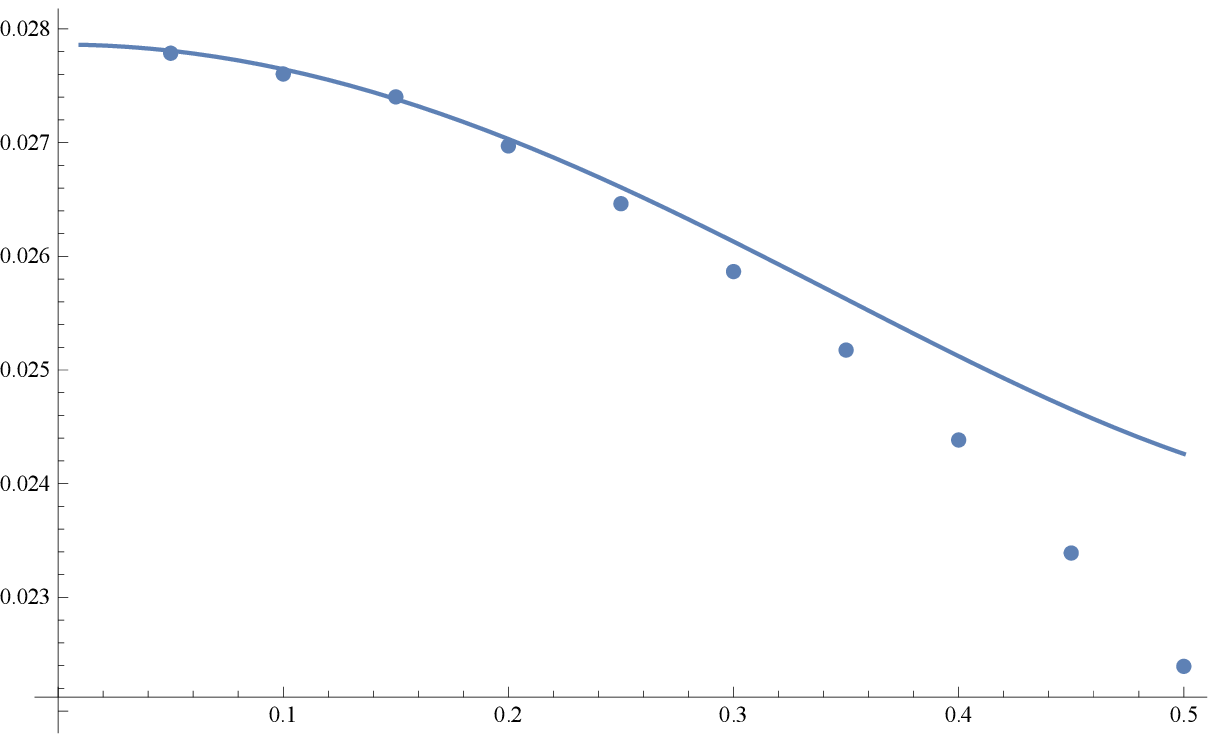}}\vfill
\caption{Comparison of the numerical results $2\sigma P_{\pm}$ and the analytic results $P_{LO}^{(S)}+P_{NLO}^{(S)}$ with $m=0.01$. We performed numerical integrals with cutoff $\int _{-100}^{100}dk \left|\int _{-\sigma}^\sigma d\tau I(\tau,k)\right|^2$. Note that when $\sigma$ is large, the analytic result is only accurate when $a$ is small.}
\label{fig:res3}
\end{figure*}

For detectors with square wave switching functions, as shown in Eq.~(\ref{eq.ap1.25}), the transition probability is given as 

\begin{equation}
\begin{split}
&P^{(S)}_{\pm}=\frac{1}{2\sigma}\left(P^{(S)}_{LO}+P^{(S)}_{NLO}\right)+\mathcal{O}(m^3),\\
&P^{(S)}_{LO}=\frac{1}{2\pi\Omega^2}\left\{-2 \text{Ci}(2 \Omega \sigma)-2 \log \left(\frac{1}{m \sigma}\right) \cos (2 \Omega \sigma)+2 \log \left(\frac{2 \Omega}{m}\right)+4 \Omega s \text{Si}(2 \Omega \sigma)-2 \pi  \Omega \sigma-2\right.\\
&\left.+\pi  \sin (2 \Omega \sigma)+2 \gamma  \cos (2 \Omega \sigma)+2 \cos (2 \Omega \sigma)+a^2\frac{2 \Omega^2 \sigma^2 \cos (2 \Omega \sigma)-4 \Omega \sigma \sin (2 \Omega \sigma)-3 \cos (2 \Omega \sigma)+3}{6 \Omega^2}\right.\\
&\left.+\frac{a^4}{180\Omega^4}\left(-2 \Omega^4 \sigma^4 \cos (2 \Omega \sigma)+8 \Omega^3 \sigma^3 \sin (2 \Omega \sigma)+18 \Omega^2 \sigma^2 \cos (2 \Omega \sigma)-24 \Omega \sigma \sin (2 \Omega \sigma)\right.\right.\\
&\left.\left.-15 \cos (2 \Omega \sigma)+15\right)+\mathcal{O}\left(\frac{\sigma^8a^6}{\Omega^2}\right)\right\},\\
&P^{(S)}_{NLO}=\frac{m^2}{4\Omega^4\pi}\left\{\log \left(\frac{64 \Omega^6}{m^6}\right)-6 \text{Ci}(2 \Omega \sigma)+\sin (2 \Omega \sigma) \left(8 \Omega \sigma \log (m \sigma)-2 \pi  \Omega ^2\sigma^2+4(2\gamma _E-1) \Omega \sigma+3 \pi \right)\right.\\
&\left.+\cos (2 \Omega \sigma) \left(\left(6-4 \Omega^2 \sigma^2\right) \log (m \sigma)-4 (\gamma _E -1) \Omega ^2 \sigma^2-4 \pi  \Omega \sigma+6 \gamma _E +5\right)+4 \Omega \sigma \text{Si}(2 \Omega \sigma)-2 \pi  \Omega \sigma-5\right\},\\
\end{split}
\end{equation}
where $\text{Ci}$ and $\text{Si}$ are cosine and sine integral functions, as defined in Eq.~(\ref{eq.ap1.26}). Likewise, we assume the mass of the scalar field to be small though nonzero. At the leading order of $a^2$, the coefficient of $a^2$ is $\left(2 (\Omega \sigma)^2-3\right) \cos (2 \Omega \sigma)-4 \Omega \sigma \sin (2 \Omega \sigma)+3$. Therefore the condition for antiUnruh effect  can be written in ``closed form" as
\begin{equation}
\begin{split}
\left(2 (\Omega \sigma)^2-3\right) \cos (2 \Omega \sigma)-4 \Omega \sigma \sin (2 \Omega \sigma)+3<0. \label{squarecond}
\end{split}
\end{equation}
The comparison of the analytic and numerical results is shown in Fig.~\ref{fig:res3}.

It can be checked easily that the condition of antiUnruh effect  for detectors with square-wave switching functions is quite different from  that for detectors with Gaussian switching functions. The antiUnruh effect  can be found not as $\Omega\sigma \to 0$ but at, for example, $\Omega \sigma =20.5\pi$. This means that antiUnruh effects  occur even when the interaction time  is long (with the energy gap fixed). Therefore our results support the argument that antiUnruh effect  are not due to non-equilibrium transient effects~\cite{Brenna:2015fga}, since the KMS condition~\cite{Kubo:1957mj,PhysRev.115.1342} is satisfied~\cite{Brenna:2015fga,Garay:2016cpf}. Furthermore, the condition Eq.~(\ref{squarecond}) depends on $\Omega \sigma$ in the form of sine and cosine function, which can be naturally expected from the Fourier transformation of the square wave function Eq.~(\ref{Frou}). In particular, this means that for some given energy gap, antiUnruh effect  can be found from time to time with the increase of $\sigma$, which is a surprising result.

\subsection{(3+1)-dimensional spacetime}
We conclude this section by displaying expressions for the transition probability of Unruh-DeWitt detectors with Gaussian switching functions in (3+1)-dimensional spacetime. As shown in Eq.~(\ref{eq.ap1.31}), the result can be obtained as
\begin{equation}
\begin{split}
&P_{\pm}^{D=3+1}=\frac{1}{2\pi\sigma^2}\left(P_a^{(0)}+
P_a^{(2)}\right)+\mathcal{O}(a^4),\\
&P_a^{(2)}=\frac{a^2 \sigma^2}{24\pi}e^{-\Omega^2\sigma^2}+\mathcal{O}(m).\\
\end{split}
\end{equation}

Note that $P_a^{(0)}$ is UV divergent; however $P_a^{(0)}\sim \mathcal{O}(a^0)$ and is therefore  of little concern to us. The dependence of $P_a^{(2)}$ on $a$ shows that when $a$ is small there is no antiUnruh effect  in the small mass limit. The same result for massless scalar field can be obtained by using a somewhat different method in ~\cite{Wu:2023glc}.

\section{Conclusion}
We obtain the analytic conditions for antiUnruh effect  in (1+1)-dimensional spacetime. The product of the detector's energy gap $\Omega$ and the interaction time $\sigma$ is the characteristic quantity in the conditions. We show that for detectors with Gaussian switching functions, the condition is $\Omega \sigma < \frac{1}{\sqrt{2}}$. However, for detectors with square wave switching functions, antiUnruh effect  could happen when $\Omega \sigma$ is large. Furthermore, for a fixed energy gap, whether antiUnruh effect  occur or not depends on the interaction time non-monotonically. Our results support the argument that antiUnruh effect  is in accordance with the KMS condition and is  therefore not a transient effect. We hope that our calculations would provide some insight on the physical nature of antiUnruh effect. Finally we show that for detectors with Gaussian switching functions there is no antiUnruh effect  in (3+1)-dimensional spacetime.

\begin{acknowledgments}
1
\end{acknowledgments}

\appendix

\section{The analytic results with small mass}
In general, the integral to be calculated can be written as
\begin{equation}
\begin{split}
&P_{\pm}=\int d^d k\left|\int _{-\infty}^{\infty}d\tau \frac{1}{\sqrt{4\pi \omega}}\chi (\tau, \sigma)\exp \left( i\Omega \tau +i\frac{\omega}{a}\sinh (a\tau ) -i\frac{k_x}{a}\left(\cosh (a\tau)-1\right) \right)\right|^2.\\
\end{split}
\label{eq.ap1.1}
\end{equation}
where $\omega \equiv  \sqrt{k^2+m^2}$, $\chi (\tau, \sigma)$ is the switching function, and $\Omega$ is defined as $\pm \Omega _0$ for $P_{\pm}$.

\subsection{\label{sec:ap1.1}The case of {D=1+1}}
It is convenient to integrate over $k$ first. The integral can be written in a somewhat symmetric form as
\begin{equation}
\begin{split}
&P_{\pm}=\int _{-\infty}^{\infty}d\tau_1\int _{-\infty}^{\infty}d\tau_2 \chi (\tau_1, \sigma)\chi (\tau_2, \sigma) e^{i\Omega (\tau _2-\tau _1)} \left(P_k(A,B)+P_k(A,-B)\right),\\
&P_k(A,B)=\int _0^{\infty}dk \frac{1}{4\pi \sqrt{m^2+k^2}} \exp \left(i\left(A\sqrt{m^2+k^2}-Bk\right)\right),\\
&A=\frac{\sinh (a\tau_2 )-\sinh (a\tau_1 )}{a},\;\;B=\frac{\cosh (a\tau_2)-\cosh (a\tau_1)}{a}.\\
\end{split}
\label{eq.ap1.2}
\end{equation}
In the case of small mass, one have
\begin{equation}
\begin{split}
&P_{k}(A,B)=\int _0^{\infty}dk \frac{1}{4\pi \sqrt{m^2+k^2}}\left(e^{i(A-B)k}+\frac{iAm^2}{2k}e^{i(A-B)k}+\mathcal{O}(\frac{Am^4}{k^3})\right).\\
\end{split}
\label{eq.ap1.3}
\end{equation}
The leading-order term of Eq.~(\ref{eq.ap1.3}) can be integrated out as
\begin{equation}
\begin{split}
&\int _0^{\infty}dk \frac{1}{\sqrt{m^2+k^2}}\exp(iC k))=-\hat{F}\left(1,\frac{C^2 m^2}{4}\right)-\frac{1}{2} i \pi  \pmb{L}_0(C m)+\log \left(-\frac{1}{2} i C m\right) (-I_0(C m))\\
&=-\log \left(-\frac{1}{2}iCm\right)-\gamma _E-i C m-\frac{C^2m^2}{4}\left(\log(-\frac{1}{2}iCm)-\gamma _E+1\right)+\mathcal{O}(m^3),\\
\end{split}
\label{eq.ap1.4}
\end{equation}
where $\gamma _E\approx 0.57721$ is the Euler constant and $I_0$ is modified Bessel function of the first kind. $\pmb{L}_0$ is modified Struve function, and $\hat{F}$ is defined as
\begin{equation}
\begin{split}
&\hat{F}(a,z)\equiv \left.\frac{d}{da'}\left(\frac{_0 F_1(a'|z)}{\Gamma (a')}\right)\right|_{a'=a},\\
\end{split}
\label{eq.ap1.5}
\end{equation}
where $_p F_q$ is the generalized hypergeometric function defined as
\begin{equation}
\begin{split}
&\;_pF_q\left(\left.\begin{array}{c}a_1,a_2,...,a_p\\b_1,b_2,...,b_q\end{array}\right|x\right)=\sum _{n=0}^{\infty} \frac{\prod_{i=1}^{p} (a_i)_n}{\prod _{j=1}^{q} (b_j)_n}\frac{x^n}{n!}.
\end{split}
\label{eq.ap1.6}
\end{equation}
Verified by numerical results, we conclude when mass is small,
\begin{equation}
\begin{split}
&\int _0^{\infty}dk \frac{1}{\sqrt{m^2+k^2}}\exp\left(i(A\sqrt{k^2+m^2} -Bk )\right)= -\log \left(-\frac{1}{2}i(A-B)m\right)-\gamma _E + \mathcal{O}(m).\\
\end{split}
\label{eq.ap1.7}
\end{equation}

Next we can  calculate the next-to-leading order term. Note that the integral can be written as
\begin{equation}
\begin{split}
&P_{k}(A,B)=P_{k}(0,B)+\int _0^A d A' \frac{\partial P_{k}(A',B)}{\partial A'},\\
\end{split}
\label{eq.ap1.8}
\end{equation}
where the first term $P_k(0,B)$ is already known in Eq.~(\ref{eq.ap1.4}) as
\begin{equation}
\begin{split}
&P_k(0,B)=-\log \left(-\frac{1}{2}iBm\right)-\gamma _E-i B m-\frac{B^2m^2}{4}\left(\log(-\frac{1}{2}iBm)-\gamma _E+1\right)+\mathcal{O}(m^3).\\
\end{split}
\label{eq.ap1.9}
\end{equation}
We define the integrand of the second term as $p(m)$
\begin{equation}
\begin{split}
&p(m)\equiv \frac{\partial P_k(A',B)}{\partial A'}=\frac{i}{4\pi}\int _0^{\infty}dk e^{i(A'\sqrt{m^2+k^2}-Bk)},
\end{split}
\label{eq.ap1.10}
\end{equation}
and similarly,
\begin{equation}
\begin{split}
&p(m)=p(0)+\frac{i}{4\pi}\int _0^m dm'\int _0^{\infty} dk\frac{\partial e^{i(A'\sqrt{m'^2+k^2}-Bk)}}{\partial m'}.\\
\end{split}
\label{eq.ap1.11}
\end{equation}
The first term can be integrated out, while the second term is
\begin{equation}
\begin{split}
&\frac{i}{4\pi}\int _0^{\infty} dk\frac{\partial e^{i(A'\sqrt{m'^2+k^2}-Bk)}}{\partial m'}=-\frac{A'm'}{4\pi} \int _0^{\infty}dk\frac{1}{\sqrt{m'^2+k^2}}
e^{i(A'\sqrt{m'^2+k^2}-Bk)}, 
\end{split}
\label{eq.ap1.12}
\end{equation}
with the leading-order term also already known in Eq.~(\ref{eq.ap1.7}). Therefore we have
\begin{equation}
\begin{split}
&p(m)=-\frac{1}{A'-B} -\frac{A'm^2}{2}\left(-\log\left(-\frac{i}{2}(A'-B)m\right)-\gamma _E+\frac{1}{2}\right)+\mathcal{O}(m^3)\\
\end{split}
\label{eq.ap1.13}
\end{equation}
and using Eqs.~(\ref{eq.ap1.2}), (\ref{eq.ap1.8} - \ref{eq.ap1.10}) and (\ref{eq.ap1.13}),
\begin{equation}
\begin{split}
&P_{\pm}=\frac{1}{4\pi}\int d\tau _1d\tau _2 \chi (\tau_1, \sigma)\chi (\tau_2, \sigma)\exp (i\Omega (\tau _2-\tau _1))\left(2\log\frac{2a}{m}-2\log\left(2i\sinh \left(\frac{a(\tau _1-\tau _2)}{2}\right)\right)-2\gamma _E\right.\\
&\left.+\frac{1}{4} m^2 (\tau_1-\tau _2)^2 ( 2 \log (i m (\tau_1-\tau _2))-2+2 \gamma _E -\log (4))\right)+\mathcal{O}(m^3).\\
\end{split}
\label{eq.ap1.14}
\end{equation}

\subsubsection{\label{sec:ap1.1.1}The Gaussian switching function}

The Gaussian switching function can be written as
\begin{equation}
\begin{split}
&\chi^{(G)} (\tau, \sigma)=\frac{1}{\sqrt{2\pi}\sigma}e^{-\frac{\tau ^2}{2\sigma^2}}.
\end{split}
\label{eq.ap1.15}
\end{equation}
Using
\begin{equation}
\begin{split}
&T=\frac{\tau _1+\tau _2}{2},\;\;t=\tau _1-\tau _2,\\
\end{split}
\label{eq.ap1.16}
\end{equation}
and integrating over $T$ first, we get
\begin{equation}
\begin{split}
&P_{\pm}^{(G)}=\frac{1}{2\pi \sigma ^2}\left(P_{LO}^{(G)}+P_{NLO}^{(G)}\right)+\mathcal{O}(m^3),\\
&P_{LO}^{(G)}=\sigma ^2e^{-\Omega ^2 \sigma ^2}\left\{\Omega ^2\sigma^2 \;_2F_2\left(\left.\begin{array}{c} 1, 1\\ \frac{3}{2}, 2\end{array}\right|\Omega^2\sigma^2\right)-\log \frac{m\sigma}{2}-\frac{\pi}{2} {\rm erfi}(\Omega \sigma)-\frac{\gamma _E}{2}\right\}+I_t,\\
&P_{NLO}^{(G)}=\frac{2m^2\sigma^4}{\sqrt{\pi}}\left(2 \Omega\sigma \left.\left(\frac{d}{dx}\; _1F_1\left(\left.\begin{array}{c} x \\ \frac{3}{2}\end{array}\right|-\Omega^2\sigma^2\right)\right) \right|_{x=2}\right.\\
&\left.-\left(\left(2 \Omega^2\sigma^2-1\right) F(\Omega \sigma)-\Omega \sigma\right) (2 \log (m \sigma)+\pi +\gamma _E -1)\right),\\
\end{split}
\label{eq.ap1.17}
\end{equation}
where $I_t$ is defined as
\begin{equation}
\begin{split}
&I_t=-\frac{\sqrt{\pi}\sigma }{\pi}\int _0^{\infty}dt \exp \left(-\frac{t^2}{4\sigma^2}\right)\cos\left(\Omega t\right)\log\left(\frac{2\sinh{\frac{at}{2}}}{at}
\right).\\
\end{split}
\label{eq.ap1.18}
\end{equation}
Considering only the case in which $a<1$, we have
\begin{equation}
\begin{split}
&\log\left(\frac{2\sinh{\frac{at}{2}}}{at}\right)=\frac{1}{24}a^2 t^2-\frac{1}{2880}a^4t^4+\frac{1}{181440}
a^6t^6+\mathcal{O}(a^8),\\
\end{split}
\label{eq.ap1.19}
\end{equation}
therefore
\begin{equation}
\begin{split}
&I_t=\frac{1}{24}I_t^1-\frac{1}{2880}I_t^2+\frac{1}{181440}I_t^3+\mathcal{O}(a^8\sigma ^{10}),\\
&I_t^n=-\frac{\sqrt{\pi}\sigma }{\pi}\int _0^{\infty}dt \exp \left(-\frac{t^2}{4\sigma^2}\right)\cos\left(\Omega t\right)a^{2n} t^{2n}\\
&=-\pi ^{\frac{1}{4}} 2^n \sigma^2 e^{-\frac{1}{2} \Omega^2 \sigma^2} \sqrt{\frac{(2 n)!}{\Omega}} (i a \sigma)^{2 n} \phi _{2n}(\Omega, \sigma)\sim \mathcal{O}(a^{2n}\sigma ^{2n+2}),\\
\end{split}
\label{eq.ap1.20}
\end{equation}
where $\phi _{2n}(\Omega \sigma)$ is the wave function of harmonic oscillator defined as
\begin{equation}
\begin{split}
&\phi _{n}(\Omega, \sigma)\equiv \frac{\left(\frac{\Omega ^2}{\pi}\right)^{\frac{1}{4}}}{\sqrt{2^n n!}}e^{-\frac{\Omega^2 \sigma ^2}{2}} H_n(\Omega \sigma),
\end{split}
\label{eq.ap1.21}
\end{equation}
and $H_n(x)$ is the Hermit polynomial.

We keep the result to order $\mathcal{O}(a^4\sigma ^6)$ and obtain
\begin{equation}
\begin{split}
&P_{LO}^{(G)}=\sigma ^2e^{-\Omega ^2 \sigma ^2}\left\{\left(\Omega ^2\sigma^2 \;_2F_2\left(\left.\begin{array}{c} 1, 1\\ \frac{3}{2}, 2\end{array}\right|\Omega^2\sigma^2\right)-\log \frac{m\sigma}{2}-\frac{\pi}{2} {\rm erfi}(\Omega \sigma)-\frac{\gamma _E}{2}\right)\right.\\
&\left.-\frac{a^2\sigma^2}{12}\left(1-2\Omega^2\sigma^2-\frac{a^2\sigma^2}{60}\left(4\Omega^4\sigma^4-12\Omega^2\sigma^2+3\right)\right)\right\}+\mathcal{O}(a^6\sigma^8).\\
\end{split}
\label{eq.ap1.22}
\end{equation}

\subsubsection{\label{sec:ap1.1.2}The square wave switching function}

The square wave switching function can be written as
\begin{equation}
\begin{split}
&\chi ^{(S)}(\tau, \sigma)=\frac{1}{2\sigma}H(\sigma - \tau)H(\sigma + \tau).
\end{split}
\label{eq.ap1.23}
\end{equation}
where $H(x)$ is the Heaviside step function.

Also using Eq.~(\ref{eq.ap1.14}) and the variable substitution in Eq.~(\ref{eq.ap1.16}), we can easily integrate $T$ out and obtain 
\begin{equation}
\begin{split}
&P^{(S)}_{\pm}=\frac{1}{\sigma}{\rm Re}\left[\frac{1}{4\pi}\int _0^{2\sigma} dt (2\sigma -t)\exp (-i\Omega t)\left(2\log\frac{2a}{m}-2\log\left(2i\sinh \left(\frac{at}{2}\right)\right)\right.\right.\\
&\left.\left.-2\gamma _E+\frac{1}{4} m^2 t^2 ( 2 \log (i m t)-2+2 \gamma _E -\log (4))\right)\right]+\mathcal{O}(m^3).\\
\end{split}
\label{eq.ap1.24}
\end{equation}
Similarly, we use the expansion in Eq.~(\ref{eq.ap1.19}) and find
\begin{equation}
\begin{split}
&P^{(S)}_{\pm}=\frac{1}{2\sigma}\left(P^{(S)}_{LO}+P^{(S)}_{NLO}\right)+\mathcal{O}(m^3),\\
&P^{(S)}_{LO}=\frac{1}{2\pi\Omega^2}\left\{-2 \text{Ci}(2 \Omega \sigma)-2 \log \left(\frac{1}{m \sigma}\right) \cos (2 \Omega \sigma)+2 \log \left(\frac{2 \Omega}{m}\right)+4 \Omega s \text{Si}(2 \Omega \sigma)-2 \pi  \Omega \sigma-2\right.\\
&\left.+\pi  \sin (2 \Omega \sigma)+2 \gamma  \cos (2 \Omega \sigma)+2 \cos (2 \Omega \sigma)+a^2\frac{2 \Omega^2 \sigma^2 \cos (2 \Omega \sigma)-4 \Omega \sigma \sin (2 \Omega \sigma)-3 \cos (2 \Omega \sigma)+3}{6 \Omega^2}\right.\\
&\left.+\frac{a^4}{180\Omega^4}\left(-2 \Omega^4 \sigma^4 \cos (2 \Omega \sigma)+8 \Omega^3 \sigma^3 \sin (2 \Omega \sigma)+18 \Omega^2 \sigma^2 \cos (2 \Omega \sigma)-24 \Omega \sigma \sin (2 \Omega \sigma)\right.\right.\\
&\left.\left.-15 \cos (2 \Omega \sigma)+15\right)+\mathcal{O}\left(\frac{\sigma^8a^6}{\Omega^2}\right)\right\},\\
&P^{(S)}_{NLO}=\frac{m^2}{4\Omega^4\pi}\left\{\log \left(\frac{64 \Omega^6}{m^6}\right)-6 \text{Ci}(2 \Omega \sigma)+\sin (2 \Omega \sigma) \left(8 \Omega \sigma \log (m \sigma)-2 \pi  \Omega ^2\sigma^2+4(2\gamma _E-1) \Omega \sigma+3 \pi \right)\right.\\
&\left.+\cos (2 \Omega \sigma) \left(\left(6-4 \Omega^2 \sigma^2\right) \log (m \sigma)-4 (\gamma _E -1) \Omega ^2 \sigma^2-4 \pi  \Omega \sigma+6 \gamma _E +5\right)+4 \Omega \sigma \text{Si}(2 \Omega \sigma)-2 \pi  \Omega \sigma-5\right\},\\
\end{split}
\label{eq.ap1.25}
\end{equation}
where $\text{Ci}$ and $\text{Si}$ are cosine and sine integral functions defined as
\begin{equation}
\begin{split}
&\text{Ci}(z)\equiv -\int _z^{\infty}dt \frac{\cos (t)}{t},\;\;\text{Si}(z)\equiv \int _0^zdt \frac{\sin (t)}{t}.
\end{split}
\label{eq.ap1.26}
\end{equation}

\subsection{\label{sec:ap1.2}The case of {D=3+1}}

In the case of $D=3+1$, the integral is UV divergent. However, we can still extract how $P_{\pm}$ depends on $a$ with small mass and small $a$. Expanding the integrand over $a$, we obtain 
\begin{equation}
\begin{split}
&\exp\left(i \frac{i\omega}{a}\sinh\left(a \tau\right)-\frac{ik_x}{a}\left(\cosh\left(a\tau\right)-1\right)\right)\\
&=e^{i\tau\omega}-\frac{1}{2}ie^{i\tau\omega}k_xt^2a-\frac{1}{24} a^2 \left(\tau^3 e^{i \tau \omega} \left(3 k_x^2 \tau-4 i \omega\right)\right)+\mathcal{O}(a^3).\\
\end{split}
\label{eq.ap1.27}
\end{equation}
We integrate over $\tau$ using Gaussian switching function, and find
\begin{equation}
\begin{split}
&I\equiv \left|\int _{-\infty}^{\infty}dte^{-\frac{t^2}{2\sigma^2}}e^{i\Omega t}\exp\left(i \frac{i\omega}{a}\sinh\left(a t\right)-\frac{ik_x}{a}\left(\cosh\left(at\right)-1\right)\right)\right|^2=I_a^{(0)}+I_a^{(2)}+\mathcal{O}(a^4),\\
&I_a^{(0)}=2\pi \sigma^2 \exp \left(-(\omega+\Omega)^2\sigma^2\right),\\
&I_a^{(2)}=\frac{1}{3} \pi  a^2 \sigma^6 e^{-(\omega+\Omega)^2\sigma^2} \left(6 k_x^2 \Omega^2 \sigma^2+12 k_x^2 \Omega \sigma^2 \omega+6 k_x^2 \sigma^2 \omega^2-3 k_x^2+2 \Omega^3 \sigma^2 \omega\right.\\
&\left.+6 \Omega^2 \sigma^2 \omega^2+6 \Omega \sigma^2 \omega^3-6 \Omega \omega+2 \sigma^2 \omega^4-6 \omega^2\right).
\end{split}
\label{eq.ap1.28}
\end{equation}
Using
\begin{equation}
\begin{split}
&\int d^dk f(|k|)k_i^2=\int d^dk f(|k|)\frac{k^2}{d},
\end{split}
\label{eq.ap1.29}
\end{equation}
we obtain
\begin{equation}
\begin{split}
&\frac{1}{16\pi^3}\int d^dk \frac{1}{\omega}I_a^{(2)}=\frac{1}{4\pi^2}\int _0^{\infty}dk \frac{k^2}{\omega}\frac{1}{3} \pi  a^2 \sigma^6 e^{-(\omega+\Omega)^2\sigma^2} \left(2 k^2 \Omega^2 \sigma^2+4k^2 \Omega \sigma^2 \omega+2 k^2 \sigma^2 \omega^2-k^2+2 \Omega^3 \sigma^2 \omega\right.\\
&\left.+6 \Omega^2 \sigma^2 \omega^2+6 \Omega \sigma^2 \omega^3-6 \Omega \omega+2 \sigma^2 \omega^4-6 \omega^2\right).
\end{split}
\label{eq.ap1.30}
\end{equation}
It is possible to obtain analytic results when $m\to 0$, that is,
\begin{equation}
\begin{split}
&P_{\pm}^{D=3+1}=\frac{1}{2\pi\sigma^2}\left(P_a^{(0)}+P_a^{(2)}\right)+\mathcal{O}(a^4)\\
&P_a^{(2)}=\frac{1}{4\pi^2}\frac{1}{3} \pi  a^2 \sigma^6\int _0^{\infty}dk k^2 e^{-(k+\Omega)^2\sigma^2} \left(8 k \Omega^2 \sigma^2+10k^2 \Omega \sigma^2 +4 k^3 \sigma^2 -7k+2 \Omega^3 \sigma^2-6 \Omega\right)+\mathcal{O}(m)\\
&=\frac{a^2 \sigma^2}{24\pi}e^{-\Omega^2\sigma^2}+\mathcal{O}(m).\\
\end{split}
\label{eq.ap1.31}
\end{equation}

%\nocite{*}
%\twocolumngrid

\begin{thebibliography}{99}
\bibitem{Unruh:1976db} W. G. Unruh, Notes on black hole evaporation, Phys. Rev. D {\bf 14}, 870 (1976).
\bibitem{Davies:1974th} P. C. W. Davies, Scalar particle production in Schwarzschild and Rindler metrics, J. Phys. A {\bf 8}, 609 (1975).
\bibitem{PhysRevD.7.2850} S. A. Fulling, Nonuniqueness of canonical field quantization in Riemannian space-time, Phys. Rev. D {\bf 7}, 2850 (1973).
\bibitem{Unruh:1983ms} W. G. Unruh and R. M. Wald, What happens when an accelerating observer detects a Rindler particle, Phys. Rev. D {\bf 29}, 1047 (1984).
\bibitem{Crispino:2007eb} L. C. B. Crispino, A. Higuchi, and G. E. A. Matsas, The Unruh effect and its applications, Rev. Mod. Phys. {\bf 80}, 787 (2008), arXiv:0710.5373 [gr-qc].
\bibitem{Brenna:2015fga} W. G. Brenna, R. B. Mann, and E. Martin-Martinez, Anti-Unruh Phenomena, Phys. Lett. B {\bf 757}, 307 (2016), arXiv:1504.02468 [quant-ph].
\bibitem{Li:2018xil} T. Li, B. Zhang, and L. You, Would quantum entanglement be increased by anti-Unruh effect?, Phys. Rev. D {\bf 97}, 045005 (2018), arXiv:1802.07886 [gr-qc].
\bibitem{Foo:2021gkl} J. Foo, R. B. Mann, and M. Zych, Entanglement amplification between superposed detectors in flat and curved spacetimes, Phys. Rev. D {\bf 103}, 065013 (2021), arXiv:2101.01912 [quant-ph].
\bibitem{Zhou:2021nyv} Y. Zhou, J. Hu, and H. Yu, Entanglement dynamics for Unruh-DeWitt detectors interacting with massive scalar fields: the Unruh and anti-Unruh effects, JHEP {\bf 09}, 088, arXiv:2105.14735 [gr-qc].
\bibitem{Chen:2021evr} Y. Chen, J. Hu, and H. Yu, Entanglement generation for uniformly accelerated atoms assisted by environmentinduced interatomic interaction and the loss of the anti-Unruh effect, Phys. Rev. D {\bf 105}, 045013 (2022), arXiv:2110.01780 [quant-ph].
\bibitem{Henderson:2019uqo} L. J. Henderson, R. A. Hennigar, R. B. Mann, A. R. H. Smith, and J. Zhang, Anti-Hawking phenomena, Phys. Lett. B {\bf 809}, 135732 (2020), arXiv:1911.02977 [gr-qc].
\bibitem{DeSouzaCampos:2020ddx} L. De Souza Campos and C. Dappiaggi, The anti-Hawking effect on a BTZ black hole with Robin boundary conditions, Phys. Lett. B {\bf 816}, 136198 (2021), arXiv:2009.07201 [hep-th].
\bibitem{deSouzaCampos:2020bnj} L. De Souza Campos and C. Dappiaggi, Ground and thermal states for the Klein-Gordon field on a massless hyperbolic black hole with applications to the anti-Hawking effect, Phys. Rev. D {\bf 103}, 025021 (2021), arXiv:2011.03812 [hep-th].
\bibitem{Robbins:2021ion} M. P. G. Robbins and R. B. Mann, Anti-Hawking phenomena around a rotating BTZ black hole, Phys. Rev. D {\bf 106}, 045018 (2022), arXiv:2107.01648 [gr-qc].
\bibitem{Conroy:2021aow} A. Conroy and P. Taylor, Response of an Unruh-DeWitt detector near an extremal black hole, Phys. Rev. D {\bf 105}, 085001 (2022), arXiv:2109.04486 [gr-qc].
\bibitem{Pan:2021nka} Y. Pan and B. Zhang, Anti-Unruh effect in the thermal background, Phys. Rev. D {\bf 104}, 125014 (2021), arXiv:2112.01889 [hep-th].
\bibitem{Barman:2021oum} S. Barman and B. R. Majhi, Radiative process of two entangled uniformly accelerated atoms in a thermal bath: a possible case of anti-Unruh event, JHEP {\bf 03}, 245, arXiv:2101.08186 [gr-qc].
\bibitem{PhysRevD.94.104048} L. J. Garay, E. Mart\'in-Mart\'inez, and J. de Ram\'on, Thermalization of particle detectors: The unruh effect and its reverse, Phys. Rev. D {\bf 94}, 104048 (2016).
\bibitem{Kubo:1957mj} R. Kubo, Statistical mechanical theory of irreversible processes. 1. General theory and simple applications in magnetic and conduction problems, J. Phys. Soc. Jap. {\bf 12}, 570 (1957).
\bibitem{PhysRev.115.1342} P. C. Martin and J. Schwinger, Theory of many-particle systems. i, Phys. Rev. {\bf 115}, 1342 (1959).
\bibitem{Garay:2016cpf} L. J. Garay, E. Martin-Martinez, and J. de Ramon, Thermalization of particle detectors: The Unruh effect and its reverse, Phys. Rev. D {\bf 94}, 104048 (2016), arXiv:1607.05287 [quant-ph].
\bibitem{Wu:2023glc} D. Wu, S.-C. Tang, and Y. Shi, Birth and death of entanglement between two accelerating Unruh-DeWitt detectors coupled with a scalar field, (2023), arXiv:2304.12126 [gr-qc].



\end{thebibliography}

\end{document}